**Hybridization gap in the heavy-fermion compound UPd$_2$Al$_3$ via quasiparticle scattering spectroscopy**

N. K. Jaggi[1,2], O. Mehio[2], M. Dwyer[2], L. H. Greene[2,a], R. E. Baumbach[3,b], P. H. Tobash[3], E. D. Bauer[3], J. D. Thompson[3], and W. K. Park[2,b,*]

[1]*Department of Physics, Illinois Wesleyan University, Bloomington, Illinois 61702-2900*

[2]*Department of Physics and Materials Research Laboratory, University of Illinois at Urbana-Champaign, Urbana, Illinois 61801, USA*

[3]*Los Alamos National Laboratory, Los Alamos, New Mexico 87545, USA*

Abstract

We present results from point-contact spectroscopy of the antiferromagnetic heavy-fermion superconductor UPd$_2$Al$_3$: conductance spectra are taken from single crystals with two major surface orientations as a function of temperature and magnetic field, and analyzed using a theory of co-tunneling into an Anderson lattice. Spectroscopic signatures are clearly identified including the distinct asymmetric double-peak structure arising from the opening of a hybridization gap when a coherent heavy Fermi liquid is formed. Both the hybridization gap, found to be 7.2 ± 0.3 meV at 4 K, and the conductance enhancement above a flat background decrease upon increasing temperature. While the hybridization gap is extrapolated to remain finite up to ~28 K, close to the temperature around which the magnetic susceptibility displays a broad peak, the conductance enhancement vanishes at ~18 K, slightly above the antiferromagnetic transition temperature ($T_N \approx 14$ K). This rapid decrease of the conductance enhancement is understood as a consequence of the junction drifting away from the ballistic regime due to increased scattering off magnons associated with the localized U 5$f$ electrons. This shows that while the hybridization gap opening is not directly associated with the antiferromagnetic ordering, its visibility in the conductance is greatly affected by the temperature-dependent magnetic excitations. Our findings are not only consistent with the 5$f$ dual-nature picture in the literature but also shed new light on the interplay between the itinerant and localized electrons in UPd$_2$Al$_3$.

*Corresponding author: wkpark@magnet.fsu.edu

[a]Present address: National High Magnetic Field Laboratory and Department of Physics, Florida State University, Tallahassee, Florida 32310, USA.

[b]Present address: National High Magnetic Field Laboratory, Florida State University, Tallahassee, Florida 32310, USA.

# I. INTRODUCTION

The 5*f*-based family of heavy-fermion compounds [1] are one among an increasing number of families of correlated electron systems, whose properties are not fully understood because strong correlations are at the core of their interesting electronic phenomena and phases (e.g., magnetic order, unconventional superconductivity). In these materials, 5*f* orbitals fall between the limits of well-localized 4*f* orbitals in the lanthanide compounds and more delocalized *d* orbitals in transition metals [2]. The periodic Anderson model containing on-site Coulomb interactions among localized electrons, and their hybridization with conduction electrons, has been solved in the mean-field approximation [1,3], and this approach has long served as a starting point for the Anderson lattice picture of these 5*f* heavy-fermion compounds where competition between Kondo coupling and the Ruderman-Kittel-Kasuya-Yosida interaction can lead to rich electronic phase diagrams.

Point-contact spectroscopy (PCS) [4,5], also referred to as quasiparticle scattering spectroscopy (QPS) [6], has contributed to our understanding of the properties of superconducting gaps, especially after the development of the Blonder-Tinkham-Klapwijk theory [7] of a superconductor/insulator/normal-metal junction with arbitrary barrier strength, which provides a robust and helpful way to identify spectral features that are directly connected to the superconducting density of states. It is only in the past dozen years, however, that significant progress has been made in measuring [6,8,9] and calculating the point-contact [10,11] or tunneling [12-14] conductance in the normal state of correlated electron systems.

In this paper, we extend our investigation of Anderson lattice heavy fermions, e.g., $URu_2Si_2$, where the PCS data were analyzed [6] with a specific and detailed theoretical model due to Maltseva, Dzero, and Coleman (MDC) [12] of co-tunneling of the injected electrons into an Anderson lattice. This co-tunneling model for these highly correlated systems is quite useful in understanding the *effective* density of electronic states, including the hybridization gap. In particular, this theory predicts and the experimental conductance spectra clearly confirm certain characteristic signatures for the lattice version of a Fano resonance [15], including a distinct asymmetric double-peak structure. Analysis based on the MDC theory enabled us to extract the hybridization gap as a function of temperature in $URu_2Si_2$ [6], where the gap was found to open well above the hidden order transition temperature [6,8], indicating that the hybridization gap is not the long-sought order parameter for its hidden order phase [16].

Conducting similar studies for related *f*-electron heavy fermions promises to provide additional insights into the nature of the correlated electronic states. In this paper, we report PCS results on single crystals of $UPd_2Al_3$ [17], in which the antiferromagnetism with a Néel temperature $T_N = 14$ K is known to be due to *localized* magnetic moments associated with some of the U 5*f* electrons [18], as opposed to more itinerant $URu_2Si_2$ where the hidden order has recently been associated with a rather exotic chiral density wave [19].

## II. PREVIOUS QPS STUDIES ON $UPd_2Al_3$

Previous PCS studies on $UPd_2Al_3$ [20-24] measured and analyzed a variety of junctions with different junction characteristics, ranging from those that were in the thermal regime, to others that were presumably in the ballistic or diffusive regime. For instance, Naidyuk *et al*. performed PCS for $URu_2Si_2$, $UPd_2Al_3$, and $UNi_2Al_3$ [23], but the zero-bias resistance of those junctions mimicked the resistivity of the bulk samples. They found asymmetric double peaks in $URu_2Si_2$, similar to what we observed [6,8], but their conductance data for $UPd_2Al_3$ was dominated by a very large and narrow single peak centered at zero bias, which is a characteristic of junctions in the thermal regime [25]. Similarly, in another detailed PCS investigation on $UPd_2Al_3$ [24], the break junction in $UPd_2Al_3$ had a resistance of 0.66 Ω, presumably rendering it in the thermal regime although signatures of the superconducting transition at $T_c$ = 1.8 K were also observed. Understandably, there were no detectable spectroscopic features in those data on $UPd_2Al_3$. Instead, hysteretic current-voltage characteristics observed below $T_N$ were described successfully in terms of the junction being distinctly in the thermal regime, resulting in local heating of 3.2 K per 1 mV [4]. While interesting in its own right, that study threw no light on the spectroscopic density of states in the normal state of this compound.

In contrast, another earlier study [21] that included measurements in the PCS configuration using a scanning tunneling microscope did observe faint but inconsistent hints of a gap-like behavior. Those experiments implied opening of a gap of about 7 − 15 meV in the density of states along the *ab*-direction, but not along the *c*-axis at 4.2 K. However, PCS data for one particular junction of resistance 70 Ω showed a very weak anomaly, which, after background subtraction, appeared to show two broad dips in the differential resistance, one at –5 mV and the other at +20 mV, suggesting an off-centered gap of approximately 25 meV. Ambiguities in these earlier PCS experiments call for a reinvestigation of the spectroscopic properties of $UPd_2Al_3$ and how these properties are influenced by the dual nature of its correlated 5*f* electrons.

## III. EXPERIMENTAL METHODS

UPd$_2$Al$_3$ single crystals were grown by the Czochralski method and oriented using a back-Laue CCD camera. The samples were characterized by electrical resistivity and magnetic susceptibility measurements. The directions of the current and the magnetic field were not defined specifically. For QPS measurements, the single crystals were cut perpendicular to the *c*- or *b*-axis, referred to as *xy*- or *xz*-cut crystal, respectively. Presumably due to thin native oxides on the crystal surface, it was difficult to get reproducible data on as-cut or even on polished crystal surfaces. The best QPS data were obtained by adding a slight ion beam etching step to the process. Appropriately cut crystals were polished to mirror-like surfaces by using diamond lapping films with particle sizes ranging from 0.5 – 9 μm and isopropyl alcohol as a lubricant. Their surfaces were then etched for two minutes using an argon ion beam (300 V, 15 mA). Immediately after the ion beam etching, the sample holder was quickly attached to the PCS rig, typically within a few minutes and subsequently inserted into a cryostat (Quantum Design, PPMS® DynaCool™). In this study, we focus only upon the non-superconducting state, so all of the presented data were taken between 4 K and 25 K. Ballistic metallic junctions were formed *in situ* at low temperature using an electrochemically polished gold tip [26] mounted on a homemade PCS rig with piezo-driven nano-positioners [27]. The differential conductance of a junction, $G(V) = dI/dV$, was measured with a standard four-probe lock-in technique as a function of temperature and applied magnetic field.

## IV. RESISTIVITY AND MAGNETIC SUSCEPTIBILITY

The upper inset of Fig. 1(a) plots the DC resistivity, $\rho(T)$, vs. temperature for one of the UPd$_2$Al$_3$ crystals used in this study. The residual resistivity ratio is 37, indicating that the crystal is of high quality, as does a sharp transition into a superconducting state at $T_c$ = 1.9 K. As shown in the lower inset of Fig. 1(a), the antiferromagnetic transition is marked by the sudden change in slope, a kink, at approximately 14 K, in good agreement with the literature [17]. The kink at $T_N$ = 14 K and the subsequent rapid decrease below $T_N$ result from a rapid decay of the scattering rate as the magnetic excitations (magnons) develop gapped spectra in the antiferromagnetic state. The measured magnetic susceptibility, $\chi(T)$, as a function of temperature is shown in Fig. 1(b), where there is an obvious and very broad peak at approximately $T_{\chi,\text{max}}$ = 35 K, far above $T_N$. Below $T_{\chi,\text{max}}$, it decreases monotonically and shows a kink at 14 K, signifying the same antiferromagnetic transition as seen in resistivity. It is known [28] that the local magnetic moments at the U sites are aligned ferromagnetically in the *ab*-plane, and successive planes along the hexagonal *c*-axis are aligned antiferromagnetically with respect to each other, as indicated in the schematic drawing, the lower inset of Fig. 1(b). The upper inset of Fig. 1(b) shows the inverse magnetic susceptibility, $\chi^{-1}(T)$, as a function of temperature. The solid line is the best linear fit to $\chi^{-1}(T)$ in the high temperature range, showing the expected Curie-Weiss behavior for the archetypical case of localized magnetic moments. The linear temperature dependence stops below $T_{\chi,nl} \sim 60$ K.

In order to further extract information on the charge transport, we fit the resistivity data in the temperature range between $T_c$ and $T_N$ to the following expression [29]:

$$\rho(T) = \rho_0 + A T^2 + C \Delta^5 e^{-\frac{\Delta}{T}} \left[ \frac{1}{5}\left(\frac{T}{\Delta}\right)^5 + \left(\frac{T}{\Delta}\right)^4 + \frac{5}{3}\left(\frac{T}{\Delta}\right)^3 \right] \tag{1}$$

This expression comes from a model that calculates the resistivity of a Fermi Liquid in the presence of magnon excitations in an antiferromagnetic state. The first two terms represent standard contributions expected for a nonmagnetic Fermi Liquid, and the third term was derived [29] as the leading contribution to the resistivity due to electron-magnon scattering in a local moment antiferromagnet, assuming that the magnon energy dispersion relation is given by $\hbar\omega_{\vec{k}} = \sqrt{\Delta^2 + S k^2}$. Here, $\Delta$ is the gap expected in the magnon spectrum of an anisotropic antiferromagnet, as is the case for UPd$_2$Al$_3$, and $S$ is the spin wave stiffness that depends upon the magnitude of the local moment and the exchange coupling.

The fitting was conducted in two steps. First, the resistivity from 2 K up to 6.2 K (where a plot of $\rho(T)$ vs. $T^2$ is visibly linear) were fitted to the first two terms, from which $\rho_0 = 8.46$ μΩcm and $A = 0.57$ μΩcmK$^{-2}$ are obtained, as shown by the solid blue line in Fig. 1(a). Then, with $\rho_0$ and $A$ constrained to these values, the parameters $C$ and $\Delta$ were varied to give the best fit to the data from 2 K up to 14 K. The resistivity data above 14 K were not included because the third term is valid only up to the Néel temperature $T_N$. The best fit, shown as the solid red line in Fig. 1(a), was obtained with $\Delta = 1.4 \pm 0.1$ meV. In an earlier study [30], the temperature dependence of the DC resistivity of a thin film UPd$_2$Al$_3$ was analyzed using a different expression and the gap was estimated to be 1.9 meV, larger than our value. Our extracted magnon excitation gap is in excellent agreement with both of the more recent inelastic neutron scattering studies [31] that found a value of $\Delta = 1.44 \pm 0.05$ meV at 2 K for the momentum transfer $\mathbf{Q_0} = (0, 0, 1/2)$ and also for $\mathbf{Q_1} = (1, 0, 1/2)$, and the other [32] that found $\Delta = 1.5 \pm 0.1$ meV at 2.5 K for $\mathbf{Q_0} = (0, 0, 1/2)$.

This value of the magnon excitation gap $\Delta$ is much smaller than the hybridization gap $\Delta_{hyb} = 7.2 \pm 0.3$ meV which we extract from the MDC analysis of our PCS conductance spectra that will be presented in the forthcoming section. $\Delta$ is a gap in the spin sector, *i.e.*, in the magnon dispersion spectrum, so can be best detected by magnetically sensitive probes such as neutron scattering, but is also detectable in electrical transport as shown above. It is fundamentally different from $\Delta_{\text{hyb}}$, which corresponds to an excitation gap in the charge sector and can be measured by PCS, as presented below.

## V. QPS DATA AND DIAGNOSTICS

Figure 2(a) displays the normalized conductance data, $G_n(V) \equiv G(V)/G(-50\ mV)$, for one particular junction on the *xz*-cut UPd$_2$Al$_3$ crystal, J8, whose level of noise and drift was low and the junction was stable over the wide temperature range from 4 – 20 K. We begin by noting that in all junctions and at all temperatures, the spectral features, *i.e.*, the asymmetric peaks in $G_n$(V), appear to ride atop a smooth and gently curved background that appears to be independent of temperature (see data in the high bias region, say, |V| > 40 mV). As the temperature increases, the intensity of this double-peak-like spectral feature decreases rapidly and, for $T \geq 20$ K, there is very little detectable trace. Thus, we carry out a further normalization of the $G_n$(V) data against the curved background. For this, $G_n$(V, 20 K) is fit to a fifth-order polynomial, $B$(V) = 1 + $b$(V), and we define the background-corrected and normalized conductance as $G_{bc}(V) \equiv G_n(V)/B(V)$. The resulting $G_{bc}$(V) spectra for the same junction are displayed in Fig. 2(b). An alternate way of background correction, *i.e.*, subtracting the curved background from the normalized conductance, $G_{bc}$(V) = $G_n$(V) – $b$(V), resulted in spectral shapes that were indistinguishable from the first method. This is because $b$(V) < 0.02 over the full range of bias voltage and $G_n$(V) itself varies by a maximum of 3 – 4% over the entire bias range. (In some junctions, the maximum variation was only 1%.)

PCS conductance data reveal spectroscopic features *only* if one can guarantee that the junction is *not* in the non-spectroscopic, namely, the thermal regime [4,5,9]. There are multiple ways to ensure this is indeed the case, as we discuss below. First, UPd$_2$Al$_3$ crystals of comparable quality, as judged by the residual resistivity ratio, reveal clear de Haas-van Alphen oscillations [33], from which the mean free path, *l*, was estimated to be 1200 Å for one of the heavy-fermion bands, ζ, and 1600 Å for another heavy band β. Using these values and assuming that ρ*l* is constant, we estimate the *l* values at 4 K to be 580 Å and 770 Å for the ζ and β band, respectively. This shows that it is, in principle, possible to make non-thermal, even ballistic junctions on these crystals. Using the standard Sharvin formula, $R_J = 16\ \rho l/3\pi d^2$ and the values of measured ρ and estimated *l*, we find *d*, the diameter of the junction whose conductance spectra are shown in Fig. 2, to be less than 400 Å at 4 K. Thus, this junction is solidly in the ballistic region (*d* < *l*) at 4 K. The same is true for other junctions whose conductance spectra are included in this paper. Of course, as the temperature increases above 4 K, the inelastic scattering cross section increases, and there is no guarantee that the junction will stay in the ballistic regime. There are two well-known signatures of a point-contact junction being in the thermal (*i.e.*, non-spectroscopic) regime, and it is necessary to ensure that conductance data don't suffer from such non-spectroscopic effects as local heating.

For a thermal junction, the voltage drop occurs entirely within the junction region via inelastic scattering processes, resulting in local heating [4]. In this case, the local junction temperature, $T_J$, becomes higher than the bath temperature, $T_{bath}$, by an amount that increases with the bias voltage, $V$, as follows: $T_J^2 = T_{bath}^2 + V^2/4L$, where $L$ is the Lorenz number. As a result, differential resistance $dV/dI$ vs. $V$ shows a strong resemblance to the bulk resistivity vs. temperature curve. For our measurements, between 4 K and 20 K, the bulk conductivity decreases monotonically with increasing temperature by more than an order of magnitude, whereas the junction conductance, $G_n(V)$ and $G_{bc}(V)$, follows a very non-monotonic dependence on $V$, as seen in Fig. 2. This point strongly indicates that our junctions are *not* in the thermal regime.

Second, if the junction were in the thermal limit, then the measured zero-bias junction resistance, $R_0$, essentially samples the bulk resistivity, $\rho(T)$. In order to scrutinize this possibility, we extract $R_0$ from the conductance data in Fig. 2 and plot it as a function of temperature in Fig. 3. Between 4 K and 20 K, $R_0$ for this particular junction *decreases* by approximately 1.7%, whereas the bulk resistivity *increases* by approximately 1600%. Namely, far from tracking the bulk resistivity, the zero-bias junction resistance is almost temperature-independent, further attesting that the junction is not in the thermal regime. The junction resistance at a bias of −50 mV, $R_{-50}$, also behaves in a similar fashion, varying by less than 2% between 4 K and 20 K. The ratio of the junction resistance at zero bias and at high bias, $R_0/R_{-50}$, increases by only about 3.5% as the temperature increases from 4 K to 20 K. This very small change is nearly three orders of magnitude less than the variation in $\rho(T)$. Finally, the absence of local heating also rules out any artifacts due to the Seebeck effect [4]. We emphasize that we include only those data sets in which these junction characteristics are reproducibly observed and that the junction dimension estimated at 4 K falls to the ballistic regime. Our careful diagnostics enables us to claim that the features seen in the conductance spectra are spectroscopic.

In UPd$_2$Al$_3$, as the temperature increases from well below $T_N$, the charge carriers undergo inelastic scattering off gapped magnons arising from the underlying antiferromagnetic structure and, thus, the inelastic scattering length, $l_{in}$, is expected to decrease, perhaps quite rapidly as the temperature approaches $T_N$. One can therefore expect that a junction that is in or close to the ballistic regime at low temperature might become diffusive at high temperature. We shall discuss this point further in section VII.

Figure 4 shows some additional representative conductance spectra $G_{bc}(V)$ under different experimental conditions. Fig. 4(a) displays the $G_{bc}(V, T)$ for 4 K $\leq$ T $\leq$ 18 K for a junction formed on an *xy*-cut crystal, *i.e.*, when electrons from the tip are injected primarily along the *c*-axis of the crystal. The main spectral features, including the overall shape, magnitude of the gap and the sign of the asymmetry, are similar to those for the *xz*-cut crystal presented in Fig. 2, but with two minor differences. First, the intensity appears to have decreased a bit more rapidly for the *xy*-cut crystal, becoming indistinguishable from the background at 18 K. Second, the extent of smearing of the peaks appears to be slightly larger than that for the *xz*-cut crystal. Figure 4(b) shows the variation of $G_{bc}(V)$ at 4 K for a junction on the same *xy*-cut crystal under an external magnetic field, $H = 0 - 9$ T, applied parallel to the *c*-axis, along with best fits to the MDC model to be discussed in the next section.

## VI. ANALYSIS OF THE QPS DATA

Our analysis adopts a similar approach as used for $URu_2Si_2$ [6], where the MDC model [12] could account for the characteristic features in the QPS data. The periodic Anderson model, in a mean-field approximation considering on-site Coulomb interaction, gives [1,3] two hybridized bands:

$$E_{k\pm} = \frac{1}{2}\left\{\varepsilon_k + \lambda \pm \sqrt{(\varepsilon_k - \lambda)^2 + 4\mathcal{V}^2}\right\} \qquad (2)$$

Here, $\lambda$ is the renormalized *f*-level and $\mathcal{V} = z^{1/2} \mathcal{V}_0$ is the renormalized hybridization matrix amplitude, where $z = 1 - n_f$ ($n_f$ being the *f*-level occupancy). According to this model, a hybridization gap opens between the two bands: a direct gap of $2\mathcal{V}$ in *k*-space and an indirect gap in the density of states given by $\Delta_{hyb} = 2\mathcal{V}^2/D$ ($2D = D_1 + D_2$ is the conduction bandwidth). Based on this hybridization picture combined with the co-tunneling mechanism, the differential conductance for tunneling into an Anderson lattice, $\left[\frac{dI}{dV}\right]_{FR}$, was derived as follows [12].

$$\left[\frac{dI}{dV}\right]_{FR} \propto \mathrm{Im}\tilde{G}^{KL}_{\psi}(eV); \quad \tilde{G}^{KL}_{\psi}(eV) = \left(1 + \frac{q_F W}{eV - \lambda}\right)^2 \ln\left[\frac{eV + D_1 - \frac{\mathcal{V}^2}{eV - \lambda}}{eV - D_2 - \frac{\mathcal{V}^2}{eV - \lambda}}\right] + \frac{2D(t_f/t_c)^2}{eV - \lambda}, \qquad (3)$$

where $-D_1$ and $D_2$ are the lower and upper conduction band edges, respectively, and the Fano parameter $q_F = t_f \mathcal{V} / t_c W$, where $t_f$ and $t_c$ are the matrix amplitudes for tunneling into the *f*-orbital and the conduction band, respectively.

Following our previous approach [6], we assume that the measured $G(V)$ is proportional to this quantity. As usual, the thermal smearing at finite temperature is taken into account by including the derivative of the Fermi function in the integration. To account for additional smearing of the conductance features due to disorder of varying kinds and also possibly due to intrinsic correlation effects [14], we introduce a single phenomenological quasiparticle broadening parameter $\Gamma$ by replacing $E$ with $E - i\Gamma$ and integrating over $E$ at each given $V$, to obtain a modified Fano resonance differential conductance, $\left[\frac{dI}{dV}\right]_{mFR}$.

A proportionality constant in the resulting $\left[\frac{dI}{dV}\right]_{mFR}$ contains two unknown parameters, the tunneling matrix element, $t_c$, from the tip into the conduction band of the Anderson lattice and the electronic density of states of the tip, $\rho_{tip}$. Further, for a given point-contact junction, there can be multiple conductance channels via nano-junctions in parallel. We can take into account the combined effects of these unknown factors by introducing a scale factor $s$ and defining a theoretical conductance, $G_{th}(V)$, as

$$G_{th}(V) = 1 + s * \left[\frac{dI}{dV}\right]_{mFR} \tag{4}$$

Before showing the outcome of our analysis, it is worth pointing out that the theoretically computed differential conductance depends upon a rather large number of parameters: $s$, $t_c$, $t_f$, $\rho_{tip}$, $\rho_c$, $W$, $\lambda$, $\mathcal{V}$, $D$, $\Gamma$, and $q_F = t_f \mathcal{V} / t_c W$, many of which are inter-related.

The conduction band density of states $\rho_c$ and its bandwidth $2D$ could be selected independently. One may set them to reasonable values as guided by band-structure calculations and treat them as constants when one varies the temperature, the magnetic field or other parameters in the model. Because of the relationship $\Delta_{hyb} = 2\mathcal{V}^2/D$, one can vary the trial value of the hybridization gap $\Delta_{hyb}$ by varying the matrix element $\mathcal{V}$, which is related to $n_f$ and $\mathcal{V}_0$ via $\mathcal{V} = (1-n_f)^{1/2} \mathcal{V}_0$. But varying $\mathcal{V}$ also changes the parameter $W = \pi \rho_c \mathcal{V}^2$, which affects the shape of the computed spectrum in ways other than simply changing the gap value. In addition, it simultaneously changes the ratio $q_F/(t_f/t_c)$ because $q_F/(t_f/t_c) = 1/(\pi \rho_c \mathcal{V})$. And, any specific pair of values $(\rho_c, \mathcal{V})$ fixes the ratio $q_F/(t_f/t_c)$, so one can at best vary only one of the two, either $t_f/t_c$ or $q_F$, independently. Because of these reasons, a brute force standard fitting to extract all of these parameters was not feasible. Instead, we systematically varied one parameter at a time, and simulated conductance spectra, $G_{th}(V)$, in order to get the best possible fits as judged by the eye and to get conservative estimates of the confidence in the corresponding fitting parameter. The uncertainty quoted for any parameter using this analysis corresponds to the change in the corresponding parameter that makes the fit visually worse than for the mean value of the corresponding parameter.

In Fig. 5, the blue curve shows the best fit for the 4 K data that one can get by adjusting all of the parameters with the exception of the quasiparticle broadening parameter $\Gamma$, which was set to zero. We show this to make two points. First and foremost, the *characteristic asymmetric double peaks* and the hybridization gap are clear, both in the data and in the theory of co-tunneling into the Anderson lattice [12]. Second, thermal smearing alone cannot account for the observed width of the peaks, and the filling of the hybridization gap needs to be attributed to a nonzero value of $\Gamma$, caused by disorder [12] and/or too short quasiparticle lifetimes due to other effects like intrinsic correlation or broken translational invariance [14]. The red curve in Fig. 5 shows our best fit when all of the parameters including $\Gamma$ were systematically varied. For this particular spectrum at 4 K, the best fit is obtained with $\Delta_{hyb} = 7.2 \pm 0.3$ meV, $\Gamma = 5.3 \pm 0.2$ meV and $s = (2.3 \pm 0.1) \times 10^{-3}$. The uncertainty in the estimated value of the parameters gets larger at higher temperatures, especially as the temperature approaches $T_N$.

Figure 6(a) shows our fit for the 10 K data when all parameters are varied. We get an excellent fit with $\Delta_{hyb} = 4.8 \pm 0.4$ meV, $\Gamma = 4.0 \pm 0.3$ meV, and $s = (8.0 \pm 0.4) \times 10^{-4}$. For the 14 K data shown in Fig. 6(c), allowing *all* of the parameters to vary independently was not helpful in extracting the parameters with confidence. In particular, if $\Gamma$ is allowed to vary independently and take on arbitrarily large values, then these data can be fit to a single broad peak. On the other hand, if we constrain $\Gamma$ to have the same value as at 10 K, the data fit to two asymmetric peaks with a hybridization gap $\Delta_{hyb} = 4.2$ meV at 14 K. Therefore, for the spectra at $T \geq 12$ K, the MDC fits were conducted with this particular constraint.

We are interested in determining the temperature at which the hybridization gap opens (or vanishes), and also the temperature at which the integrated area under the peaks goes to zero. Because i) uncertainty in the extracted parameters is large (as discussed above), ii) the MDC analysis is computationally intensive, and iii) parameters in the MDC model affect the spectra in highly correlated ways, we have conducted a simpler analysis of the data to corroborate that these two extracted temperatures are reproducible. For this, we have carried out a straightforward nonlinear least-squares fit of the spectroscopic data to a sum of two Lorentzian peaks of equal (but variable) half-width-at-half-maximum (HWHM), $\gamma_{HWHM}$, whose positions ($V_1$, $V_2$) and heights ($h_1$, $h_2$) are completely free parameters. In general, the area under a Lorentzian peak of height $h$ is given by $\pi \gamma_{HWHM} h$. This approach produced good fits for all temperatures up to 18 K, and a typical fit using this method is shown in Fig. 6(b) for the 10 K data. The peak-separation, $V_1 - V_2$, is as large as about twice $\Delta_{hyb}$.

The hybridization gap and the peak separation are plotted together in Fig. 7(a). It is clear that $\Delta_{\mathrm{hyb}}$ decreases approximately linearly with increasing temperature between 4 K and 16 K. Based solely upon our data, we cannot rule out the possibility of a discontinuous drop in $\Delta_{\mathrm{hyb}}$ at some temperature above 16 K, but a recent theoretical analysis [34] predicts that the hybridization gap in metallic heavy fermion systems decreases smoothly with increasing temperature. Therefore, we include in the figure a linear fit to the temperature dependence of $\Delta_{\mathrm{hyb}}$, which suggests that $\Delta_{\mathrm{hyb}}$ remains finite up to approximately 28 K. This shows that the hybridization gap opening temperature in UPd$_2$Al$_3$, $T_{\mathrm{hyb}}$, is much higher than the Néel temperature, namely, $T_{\mathrm{hyb}} \gg T_{\mathrm{N}}$, quite similarly to the case of URu$_2$Si$_2$ [6,8]. The peak-separation (V$_1$ − V$_2$), obtained from the double-Lorentzian fit, also shows a similar temperature dependence to that of $\Delta_{\mathrm{hyb}}$.

We quantify the net area under the conductance curve $G_{bc}$(V) by a straightforward numerical integration of $G_{bc}$(V) − 1 over the applied bias voltage range. As $G_{bc}$(V) is dimensionless, this area has units of volts. By multiplying it with the junction conductance (in unit of Ω$^{-1}$), we can compute *excess* junction current, which is a measure of additional current added on top of the ohmic current for a simple metal. Certainly, this excess current originates from the non-trivial density of states in UPd$_2$Al$_3$ as the hybridization gap opens. The temperature dependence of this excess junction current is shown in Fig. 7(b). Noticeably, it decreases with temperature more rapidly than $\Delta_{\mathrm{hyb}}$ and goes to zero between 18 and 20 K. The same quantity, computed from the area under the two Lorentzian peaks, $\pi\gamma_{\mathrm{HWHM}}(h_1 + h_2)$, is also plotted in Fig. 7(b). Clearly, both methods give similar results, ensuring that the excess current indeed disappears at a temperature distinct from $T_{\mathrm{hyb}}$. In order to facilitate the discussion in the next section, we include in Fig. 7(b) the bulk conductivity data along with its fit to the model for scattering off gapped magnons that is essentially the same as shown in Fig. 1(a).

Going back to Fig. 4(b) briefly, where the magnetic field dependent data on an *xy*-cut crystal are plotted, the solid lines are fits to the MDC model as described above. As the magnetic field was varied from 0 T to 9 T, $\Delta_{\mathrm{hyb}}$ and $\Gamma$ remained essentially unchanged within their respective uncertainties, at $\Delta_{\mathrm{hyb}} \approx$ 7.0 meV and $\Gamma \approx 5.6$ meV.

## VII. DISCUSSION

As already described previously, we have observed spectroscopic signatures in our QPS measurements on UPd$_2$Al$_3$ including the distinct *asymmetric double-peak* structure and our analysis based on the MDC model shows that a mechanism of co-tunneling into an Anderson lattice can explain those features successfully.

In our MDC analysis of the conductance spectra on UPd$_2$Al$_3$, some parameters are not well determined. As an example, the sign of the Fano parameter, $q_F$, can be determined reliably but its magnitude cannot be determined uniquely because $q_F = t_f \mathcal{V} / t_c W$ and $W = \pi \rho_c \mathcal{V}^2$ and varying these parameters affects the computed conductance spectra in highly correlated ways. However, the magnitude of the hybridization gap $\Delta_{hyb}$ and the quasiparticle broadening parameter $\Gamma$ can be extracted with some confidence. In UPd$_2$Al$_3$, $\Delta_{hyb}$ is determined to be 7.2 meV ± 0.3 meV at 4 K for the *xz*-cut crystal. It is interesting to note that the sign of $q_F$ is opposite to that in URu$_2$Si$_2$, whose origin remains to be investigated further. The extracted value of $\Gamma$ at 4 K is somewhat large: 5.3 ± 0.2 meV for the *xz*-cut crystal and 5.6 ± 0.2 meV for the *xy*-cut one, more than 70% of $\Delta_{hyb}$. Because several different mechanisms can contribute to $\Gamma$ including both intrinsic, such as strong correlations [14] and scattering from weakly or unhybridized 5*f* moments, and extrinsic effects, such as lattice disorder [12], $\Gamma$ is introduced just as a phenomenological fitting parameter to account for any smearing effects other than the thermal smearing.

We now discuss the temperature dependence of the excess junction current presented in Fig. 7(b). First, we recall that, depending upon the relative ratio of the junction size, *d*, and the elastic and inelastic electron scattering lengths ($l_{el}$ and $l_{in}$), point contact junctions can be in one of three regimes [4,5]. In the extreme thermal or Maxwell limit, $d \gg l_{in}$, the excess energy of injected electrons is dissipated within the junction area, causing accumulation of heat, and, thus, an increase in the junction temperature, and the *G*(V) spectra do not contain any spectroscopic information about the density of states in the material under study. In section V, we already provided evidence that our junctions are located far away from the thermal limit. On the other extreme, called ballistic or Sharvin limit, the junction dimension, $d \ll l_{el}, l_{in}$, and for such junctions, *G*(V) can pick up spectroscopic information on the bulk electronic states.

Between these two extremes lies the diffusive regime, where $l_{el} < d < \sqrt{l_{el} l_{in}/3}$ [4]. In this intermediate regime, spectral information is preserved but the measured conductance spectrum *G*(V) has a weaker intensity. It is expected that as $l_{in}$ decreases a larger fraction of injected electrons may undergo inelastic scattering inside the junction, meaning that they lose some energy before reaching the other side of the junction, and this fraction cannot contribute to the spectral component of the conductance. So, one can expect that the intensity of the spectral component of *G*(V) decreases as $l_{in}$ decreases. This picture can qualitatively explain the temperature dependence of the conductance and also the excess current shown in Fig. 7(b). As temperature increases, the rate of scattering off gapped magnons increases. This results in a decrease of the inelastic scattering length, $l_{in}$, which is responsible for both the decrease in conductivity, and also, as discussed above, for the reduction in the intensity of the spectral component of $G$(V), namely, the excess current. In this picture, the gapped magnons arising from the antiferromagnetically ordered structure have no effect on the temperature dependence of the hybridization gap because this gap opens between the hybridized (itinerant) bands and, thus, should be independent of the ordering of local moments. On the other hand, the increasing rate of inelastic scattering off gapped magnons directly affects the intensity, or the excess current, of the observed asymmetric conductance peaks.

We compare our results with the reports in the literature. First, an earlier study of the optical properties in UPd$_2$Al$_3$ [30] analyzed a broad structure in the temperature-dependent conductivity to infer that a hybridization gap of ~10 meV opens below a coherence temperature $T^*$ ~ 50 K. This is a bit larger than our value of $\Delta_{hyb}$ = 7.2 ± 0.3 meV at 4 K. Extrapolating the $\Delta_{hyb}$(T) data extracted in our analysis gives a zero temperature value of $\Delta_{hyb}$(0) ≈ 8 meV. From inspection of features in the optical conductivity data [30], it appears that the data are equally consistent with a gap opening temperature anywhere between 30 K and 60 K. The hybridization gap opening temperature found in our QPS study is ~28 K, close to the lower edge of their temperature range. Second, the monotonic temperature dependence of our extracted $\Delta_{hyb}$ is consistent with a theoretical calculation [34], in which the renormalized hybridization matrix element $\mathcal{V}$ and, therefore, $\Delta_{hyb}$ are shown to decrease monotonically with temperature (see Fig. 2 in Ref. 34). It remains to be investigated further what causes the discrepancy in the gap value (~50 meV) as well as the gap opening temperature (~65 K) extracted from those angle-resolved photoemission spectroscopy (ARPES) [34] measurements and our current QPS analysis.

Summarizing our observations of the hybridization gap in UPd$_2$Al$_3$, its opening temperature is much higher than $T_N$; its magnitude is essentially the same for both the *xz*-cut and the *xy*-cut crystals; and within measurement uncertainties it is independent of magnetic fields up to 9 T applied parallel to the current injection. These results are consistent with a qualitative picture for the underlying physics in UPd$_2$Al$_3$ in which one subsystem of itinerant U 5*f* electrons (with a zero temperature hybridization gap $\Delta_{hyb}$(0) ≈ 8 meV) scatter off the other subsystem of U 5*f* electrons that have strongly localized magnetic moments on the U sites. At temperatures well above $T_{\chi,max}$, there is no distinction between these two subsystems, but the onset of a hybridization gap, which corresponds closely to $T_{\chi,max}$, signals the incorporation of a fraction of U's 5*f* electrons into a heavy hybridized band. The remaining weakly or unhybridized fraction of 5*f* electrons order antiferromagnetically below 14 K. Magnetic excitations in the antiferromagnetically ordered state strongly scatter electrons in the heavy hybridized bands and qualitatively influence the hybridization gap features observed in QPS data as well as the bulk transport itself, *i.e.*, resistivity. This is reminiscent of the dual-nature picture for the U 5*f* electrons that was proposed following some early high-resolution photoemission studies [35] and refined in later theoretical investigations [36,37]. In this picture, some of the U 5*f* electrons are treated as itinerant and are described by the periodic Anderson model along with a hybridization gap, and the remaining U 5*f* electrons form a subsystem which carries localized magnetic moments on the U sites. Thus, our picture described above is not only consistent with the dual-nature scenarios proposed in the literature [35-37] but also sheds new light on the interplay among the U 5*f* electrons in $UPd_2Al_3$, some of which become itinerant and the others remain localized as the temperature is lowered.

## VIII. CONCLUSIONS

Our QPS study on $UPd_2Al_3$ detects the hybridization gap and novel Fano resonance, signified by asymmetric double peaks, as predicted for tunneling into an Anderson lattice [12] and also seen in $URu_2Si_2$ [6,8]. This gap, of approximately 7.2 meV at 4 K, opens at $T_{hyb} \sim 28$ K $\gg T_N$. The intensity of the asymmetric double peaks decreases as $T_N$ is approached from below. This is a consequence of the increasing scattering of electrons off magnons that causes a decrease in the inelastic scattering length, which in turn results in decreasing intensity of the spectroscopic features as the ballistic junction is tuned into the diffusive regime. The picture that emerges from our study is consistent with the dual-nature scenarios [35-37] that have been invoked for the multiple 5*f* electrons in $UPd_2Al_3$.

## ACKNOWLEDGMENTS

This material is based upon work supported by the U.S. National Science Foundation, Division of Materials Research, under Award No. 12-06766 and carried out in part in the Materials Research Laboratory Central Research Facilities at the University of Illinois at Urbana-Champaign. Work at Los Alamos National Laboratory was performed under the auspices of the U.S. Department of Energy, Office of Basic Energy Sciences, Division of Materials Sciences and Engineering. NKJ thanks Illinois Wesleyan University for a sabbatical leave during which the collaboration with the group at the University of Illinois at Urbana-Champaign was first initiated and later led to this study.

**APPENDIX: ON THE POSSIBILE SIGNATURES DUE TO CEF EXCITATIONS**

In this section, we briefly discuss a possibility that the features in our conductance spectra might originate from the crystal electric field (CEF) levels in $UPd_2Al_3$ rather than from the hybridization as claimed in the main text. This is because a CEF scheme has been successfully adopted to explain some of the properties in this compound, in particular, magnetic susceptibility [38]. Also, Krimmel et al. [39] reported results from inelastic neutron scattering experiments, interpreting the two peaks (at 3 and 7 meV), albeit not pronounced, as originating from excitations between the ground state and excited CEF levels.

The observation of CEF excitations in QPS measurements has been reported for some other materials in the literature [4]. In all such cases, they appear as pronounced peaks in the second derivative of the voltage-current characteristics, namely, $d^2V/dI^2$, originating from inelastic scattering of injected electrons. Because we did not directly measure this quantity from point-contact junctions on $UPd_2Al_3$ and because we took $dI/dV$ instead of $dV/dI$, we deduce a $d^2V/dI^2$ vs. V spectrum using the following relationship, $d^2V/dI^2 = -(dI/dV)^{-3} \times d^2I/dV^2$, along with numerical differentiation of the conductance data. The resultant second derivative spectrum extracted from the same 4 K data as shown in Fig. 5 is displayed in Fig. 8. It is clear that there is only one broad peak around 11 mV, distinct from both of the peak locations observed in the inelastic neutron scattering data [39]. Therefore, we rule out the possibility that the CEF levels play a major role in shaping our conductance data. In addition, the CEF levels in $UPd_2Al_3$ are unlikely to show such strong temperature dependence well below $T_N$ as seen in the extracted hybridization gap.

Speculating on the reason why no noticeable features due to the CEF excitations are observed in our QPS data, we note that the 5$f$ electrons in $UPd_2Al_3$ seem to undergo intriguing transformations in their states: Some become itinerant heavy electrons via hybridization process, whereas others remain localized and get ordered magnetically upon lowering the temperature. This situation is much more complicated than other cases in which CEF excitations have been clearly observed. For instance, in $PrNi_5$ [40], the magnetic $Pr^{3+}$ ions don't go through magnetic ordering nor hybridize with conduction electrons upon cooling and, thus, the CEF levels remain well-defined. We also note that it is desirable to obtain better-resolved neutron scattering data to unambiguously determine the CEF levels in $UPd_2Al_3$ [39] for further explorations of this topic.

REFERENCES

[1] P. Coleman, in *Handbook of magnetism and advanced magnetic materials*, edited by H. Kronmuller and S. Parkin (Wiley, New York, 2007), Vol. 1 *Fundamentals and Theory*, pp. 95-148.

[2] A. C. Hewson, *The Kondo Problem to Heavy Fermions* (Cambridge University Press, 1993).

[3] D. M. Newns and N. Read, Adv. Phys. **36**, 799 (1987).

[4] Y. G. Naidyuk and I. K. Yanson, *Point-contact spectroscopy* (Springer Science+Business Media, Inc., New York, 2005), Vol. 145.

[5] W. K. Park and L. H. Greene, J. Phys.: Condens. Matter **21**, 103203 (2009).

[6] W. K. Park, P. H. Tobash, F. Ronning, E. D. Bauer, J. L. Sarrao, J. D. Thompson, and L. H. Greene, Phys. Rev. Lett. **108**, 246403 (2012).

[7] G. E. Blonder, M. Tinkham, and T. M. Klapwijk, Phys. Rev. B **25**, 4515 (1982).

[8] W. K. Park, S. M. Narasiwodeyar, E. D. Bauer, P. H. Tobash, R. E. Baumbach, F. Ronning, J. L. Sarrao, J. D. Thompson, and L. H. Greene, Philos. Mag. **94**, 3737 (2014).

[9] W. K. Park, S. Narasiwodeyar, M. Dwyer, P. C. Canfield, and L. H. Greene, arXiv:1411.7073.

[10] M. Fogelström, W. K. Park, L. H. Greene, G. Goll, and M. J. Graf, Phys. Rev. B **82**, 014527 (2010).

[11] W. C. Lee, W. K. Park, H. Z. Arham, L. H. Greene, and P. Phillips, Proc. Natl. Acad. Sci. U. S. A. **112**, 651 (2015).

[12] M. Maltseva, M. Dzero, and P. Coleman, Phys. Rev. Lett. **103**, 206402 (2009).

[13] J. Figgins and D. K. Morr, Phys. Rev. Lett. **104**, 187202 (2010).

[14] P. Wölfle, Y. Dubi, and A. V. Balatsky, Phys. Rev. Lett. **105**, 246401 (2010).

[15] U. Fano, Phys. Rev. **124**, 1866 (1961).

[16] J. A. Mydosh and P. M. Oppeneer, Rev. Mod. Phys. **83**, 1301 (2011).

[17] C. Geibel *et al.*, Z. Phys. B **84**, 1 (1991).

[18] A. Krimmel, P. Fischer, B. Roessli, H. Maletta, C. Geibel, C. Schank, A. Grauel, A. Loidl, and F. Steglich, Z. Phys. B **86**, 161 (1992).

[19] H. H. Kung, R. E. Baumbach, E. D. Bauer, V. K. Thorsmolle, W. L. Zhang, K. Haule, J. A. Mydosh, and G. Blumberg, Science **347**, 1339 (2015).

[20] J. Aarts, A. P. Volodin, A. A. Menovsky, G. J. Nieuwenhuys, and J. A. Mydosh, Europhys. Lett. **26**, 203 (1994).

[21] A. P. Volodin, J. Aarts, A. A. Menovsky, G. J. Nieuwenhuys, and J. A. Mydosh, Physica B **194**, 2033 (1994).

[22] O. E. Kvitnitskaya, Y. G. Naidyuk, A. Nowack, K. Gloos, C. Geibel, A. G. M. Jansen, and P. Wyder, Physica B **259-61**, 638 (1999).

[23] Y. G. Naidyuk, O. E. Kvitnitskaya, A. G. M. Jansen, P. Wyder, C. Geibel, and A. A. Menovsky, Low Temp. Phys. **27**, 493 (2001).

[24] Y. G. Naidyuk, K. Gloos, I. K. Yanson, and N. K. Sato, J. Phys.: Condens. Matter **16**, 3433 (2004).

[25] Y. G. Naidyuk and I. K. Yanson, J. Phys.: Condens. Matter **10**, 8905 (1998).

[26] S. Narasiwodeyar, M. Dwyer, M. Liu, W. K. Park, and L. H. Greene, Rev. Sci. Instrum. **86**, 033903 (2015).

[27] M. Tortello, W. K. Park, C. O. Ascencio, P. Saraf, and L. H. Greene, Rev. Sci. Instrum. **87**, 063903 (2016).

[28] A. Hiess *et al.*, J. Phys.: Condens. Matter **18**, R437 (2006).

[29] E. Jobiliong, J. S. Brooks, E. S. Choi, H. Lee, and Z. Fisk, Phys. Rev. B **72**, 104428 (2005).

[30] M. Dressel, N. Kasper, K. Petukhov, D. N. Peligrad, B. Gorshunov, M. Jourdan, M. Huth, and H. Adrian, Phys. Rev. B **66**, 035110 (2002).

[31] E. Blackburn, A. Hiess, N. Bernhoeft, and G. H. Lander, Phys. Rev. B **74**, 024406 (2006).

[32] N. K. Sato, N. Aso, K. Miyake, R. Shiina, P. Thalmeier, G. Varelogiannis, C. Geibel, F. Steglich, P. Fulde, and T. Komatsubara, Nature **410**, 340 (2001).

[33] Y. Inada, H. Yamagami, Y. Haga, K. Sakurai, Y. Tokiwa, T. Honma, E. Yamamoto, Y. Onuki, and T. Yanagisawa, J. Phys. Soc. Jpn. **68**, 3643 (1999).

[34] X. D. Yang, P. S. Riseborough, and T. Durakiewicz, J. Phys.: Condens. Matter **23**, 094211 (2011).

[35] T. Takahashi, N. Sato, T. Yokoya, A. Chainani, T. Morimoto, and T. Komatsubara, J. Phys. Soc. Jpn. **65**, 156 (1996).

[36] P. Thalmeier, Eur. Phys. J. B **27**, 29 (2002).

[37] G. Zwicknagl and P. Fulde, J. Phys.: Condens. Matter **15**, S1911 (2003).

[38] A. Grauel *et al.*, Phys. Rev. B **46**, 5818 (1992).

[39] A. Krimmel, A. Loidl, R. Eccleston, C. Geibel, and F. Steglich, J. Phys.: Condens. Matter **8**, 1677 (1996).

[40] M. Reiffers, Y. G. Naidyuk, A. G. M. Jansen, P. Wyder, I. K. Yanson, D. Gignoux, and D. Schmitt, Phys. Rev. Lett. **62**, 1560 (1989).

FIGURE CAPTIONS

Figure 1. (a) DC resistivity in the temperature range of $T_c < T < T_N$ of a single crystalline $UPd_2Al_3$ used in the QPS measurement. The red solid line is the best fit, as described in the text, to the expression $\rho(T) = \rho_0 + A T^2 + C \Delta^5 e^{-\frac{\Delta}{T}} \left\{ \frac{1}{5}\left(\frac{T}{\Delta}\right)^5 + \left(\frac{T}{\Delta}\right)^4 + \frac{5}{3}\left(\frac{T}{\Delta}\right)^3 \right\}$, with $\Delta = 1.4 \pm 0.1$ meV. The upper inset shows $\rho(T)$ up to 300 K with $T_{\rho,max}$ for the resistance maximum being approximately 82 K. The lower inset is a zoomed view of $\rho(T)$ below 18 K showing both the antiferromagnetic transition (kink) at $T_N = 14$ K and the superconducting transition (jump) at $T_c = 1.9$ K. (b) Magnetic susceptibility $\chi(T)$ of the same $UPd_2Al_3$ crystal, showing that $\chi(T)$ peaks at approximately $T_{\chi, max} = 35$ K, well above the antiferromagnetic transition at $T_N = 14$ K. The upper inset shows the temperature dependence of the inverse susceptibility $\chi^{-1}$, with the solid red line representing the best linear fit in the high temperature range. $\chi^{-1}$ starts deviating from this straight line around $T_{\chi,nl} \approx 60$ K. The lower inset is a schematic of the hexagonal crystal structure of $UPd_2Al_3$ including the antiferromagnetic alignment of the local moments below $T_N$.

Figure 2. (a) Representative normalized differential conductance, $G_n(V) \equiv G(V) / G(-50mV)$, curves for the $xz$-cut $UPd_2Al_3$ crystal at various temperatures from 4 K to 20 K, shifted vertically for clarity. (b) Background corrected $G_{bc}(V)$ for the same set of conductance spectra. $G_{bc}(V) \equiv G_n(V) / B(V)$, where B(V) is a fifth order polynomial fit to the residual smooth curved background conductance that was observed above 20 K.

Figure 3. Comparison of the temperature dependence of the junction resistance $(R_J)$ at zero bias, $R_0$, and at −50 mV, $R_{-50}$, with that of the bulk resistivity, $\rho(T)$, as a function of temperature. With the temperature increasing from 4 K to 20 K, $R_0$ and $R_{-50}$ remain nearly unchanged, decreasing by less than 2%, whereas, over the same temperature range the bulk resistivity increases by approximately 1600%, as also shown. The ratio $R_0/R_{-50}$ remains close to unity, increasing with temperature by approximately 3.5%. .

Figure 4. (a) Temperature-dependent set of $G_{bc}(V)$ spectra for a point-contact junction on an $xy$-cut crystal (current injection along the $c$-axis). (b) Magnetic field dependence of $G_{bc}(V)$ at 4 K for the same junction. The solid lines are the best fits to the MDC model. The fitting parameters show nearly no field dependence: $\Delta_{hyb} = 7.0 \pm 0.4$ meV, $\Gamma = 5.7 \pm 0.3$ meV, and $s = (1.7 \pm 0.1) \times 10^{-3}$.

Figure 5. $G_{bc}(V)$ data at 4 K for the $xz$-cut crystal along with fits to the MDC model. The blue line indicates the calculation when the quasiparticle broadening parameter is set to zero, $\Gamma = 0$. The red line is the best fit obtained when $\Gamma$ is varied freely, corresponding to $\Delta_{hyb}$ = 7.2 ± 0.3 meV, $\Gamma$ = 5.3 ± 0.2 meV, and $s$ = (2.3 ± 0.1) × $10^{-3}$.

Figure 6. (a) Best fit of the 10 K data using the MDC model, with $\Delta_{hyb}$ = 4.8 meV and $\Gamma$ = 4.0 meV. (b) Fit of the same 10 K data to a sum of two Lorentzian peaks, giving the peak separation, $V_1 - V_2$ = 9.3 meV. (c) For temperature above 10 K, the peak intensity gets smaller and the data are so noisy that varying $\Delta_{hyb}$ and $\Gamma$ independently in the MDC model was not meaningful. The plotted data were taken at 14 K. The red solid line is a fit to the MDC model where $\Gamma$ was constrained to have the same value as at 10 K.

Figure 7. (a) Temperature dependence of the hybridization gap (blue triangles) and the peak separation (red circles). The dashed straight lines are linear fits. (b) Temperature dependence of the excess junction current, which is proportional to the net area between a $G_{bc}(V)$ curve and the flat background line, obtained using two methods: by a straightforward numerical integration of $G_{bc}(V) - 1$ (black triangles), and by evaluating $\pi\gamma_{HWHM}(h_1 + h_2)$ where $\gamma_{HWHM}$, $h_1$ and $h_2$ are the best fit parameters from the double-Lorentzian analysis (red squares). For comparison, the bulk conductivity (pink squares) and its fit (blue solid line) to the model from Fig. 1(a) are also plotted.

Figure 8. Second derivative spectrum, $d^2V/dI^2$ vs. V, deduced from the conductance data at 4 K shown in Fig. 5 using the relationship $d^2V/dI^2 = -(dI/dV)^{-3} \times d^2I/dV^2$ along with numerical differentiation of the conductance data.

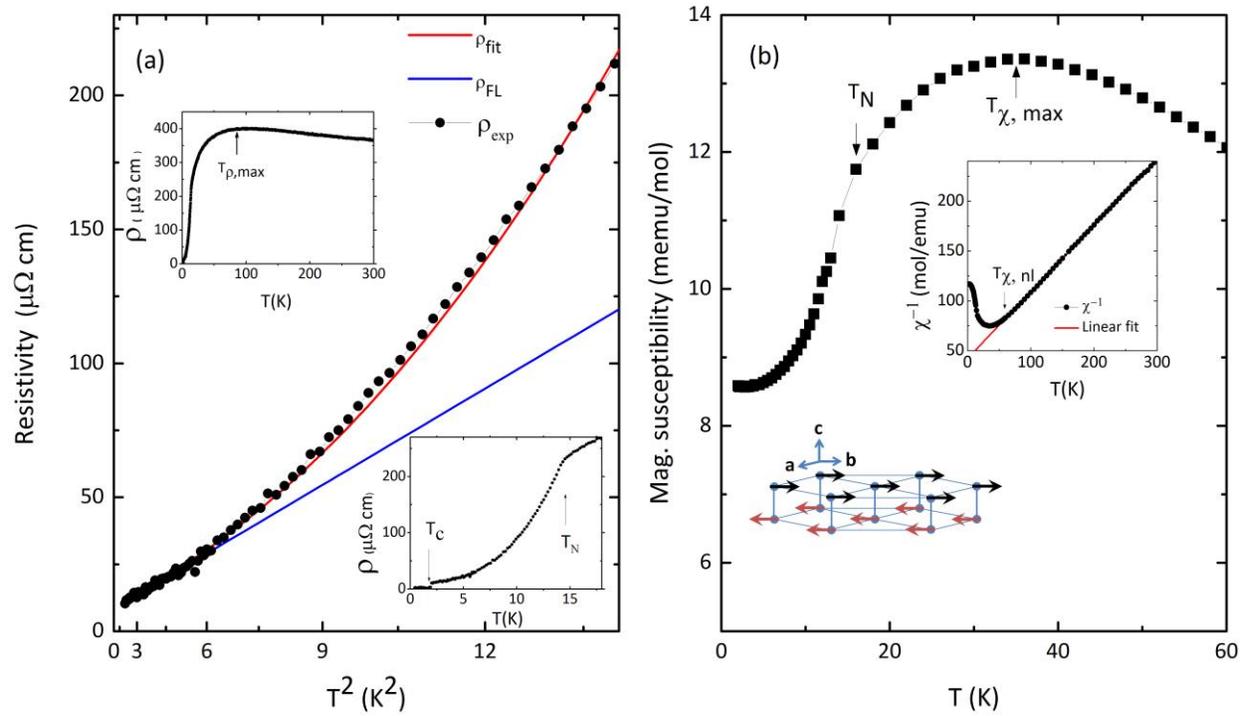

Figure 1

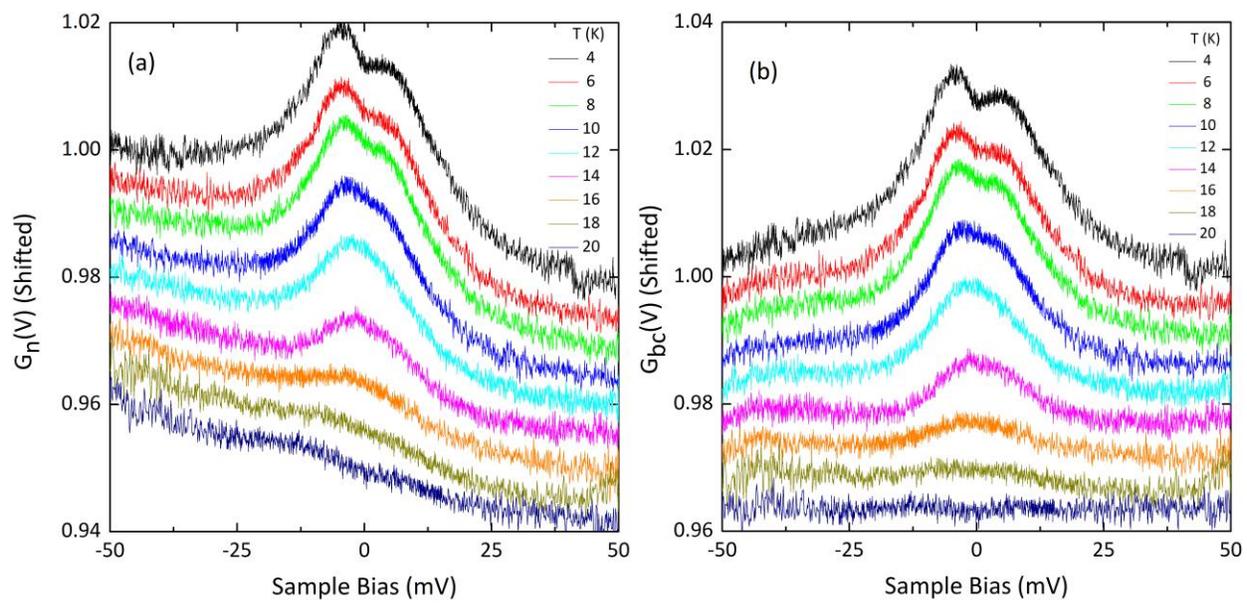

Figure 2

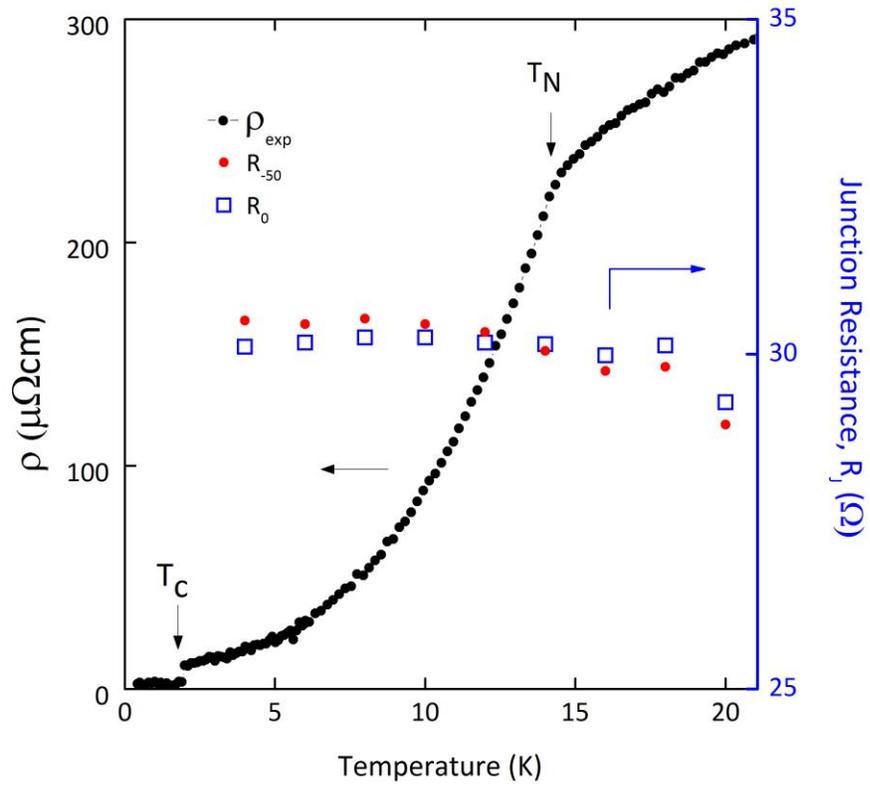

Figure 3

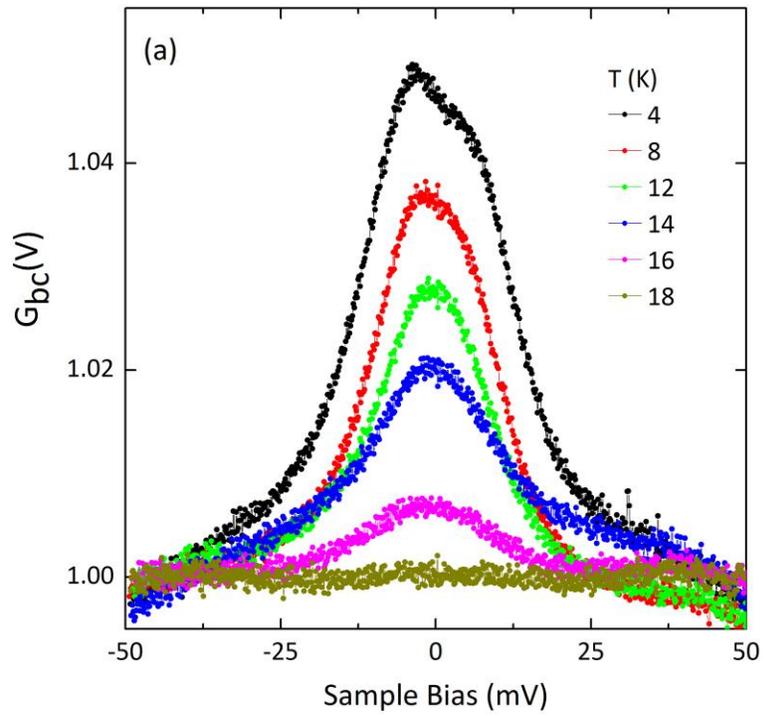
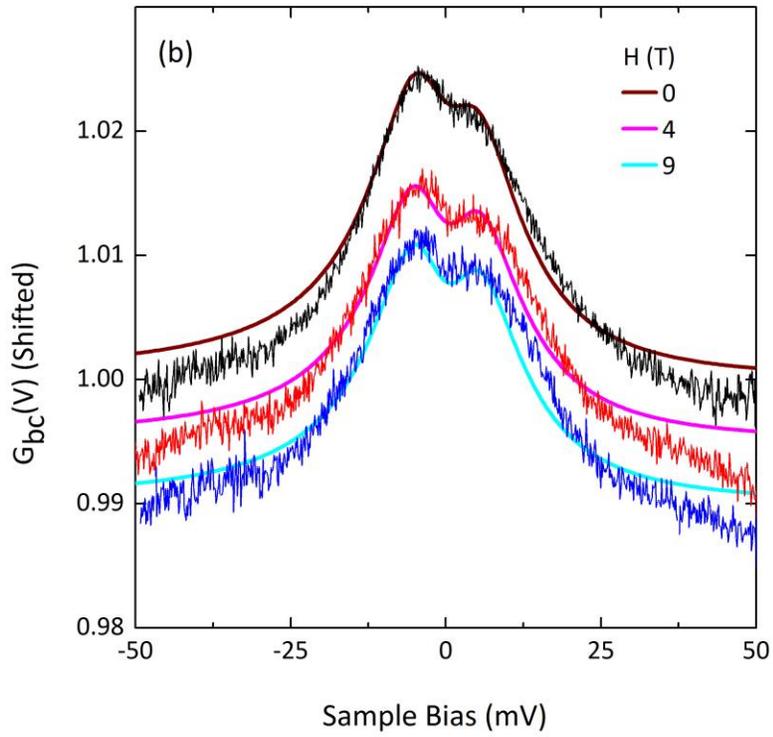

Figure 4

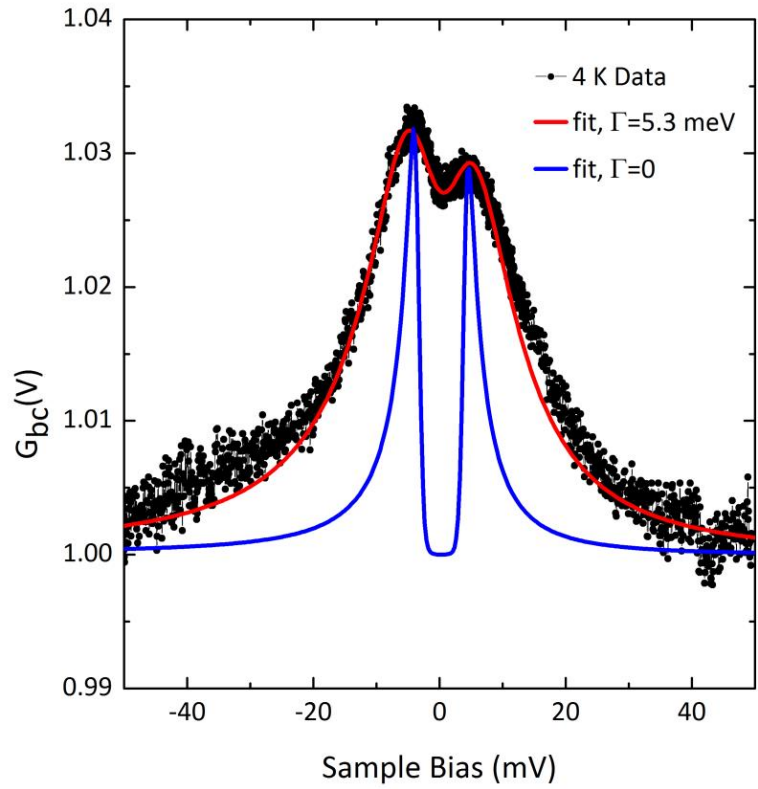

Figure 5

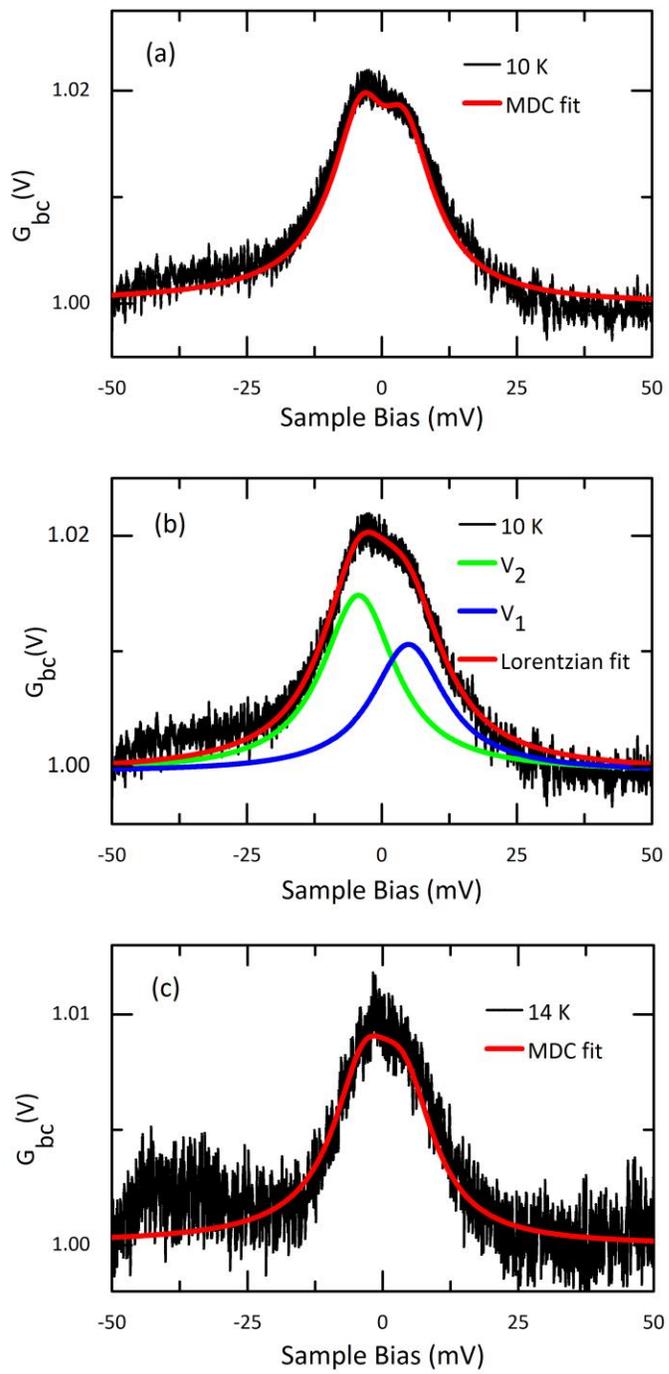

Figure 6

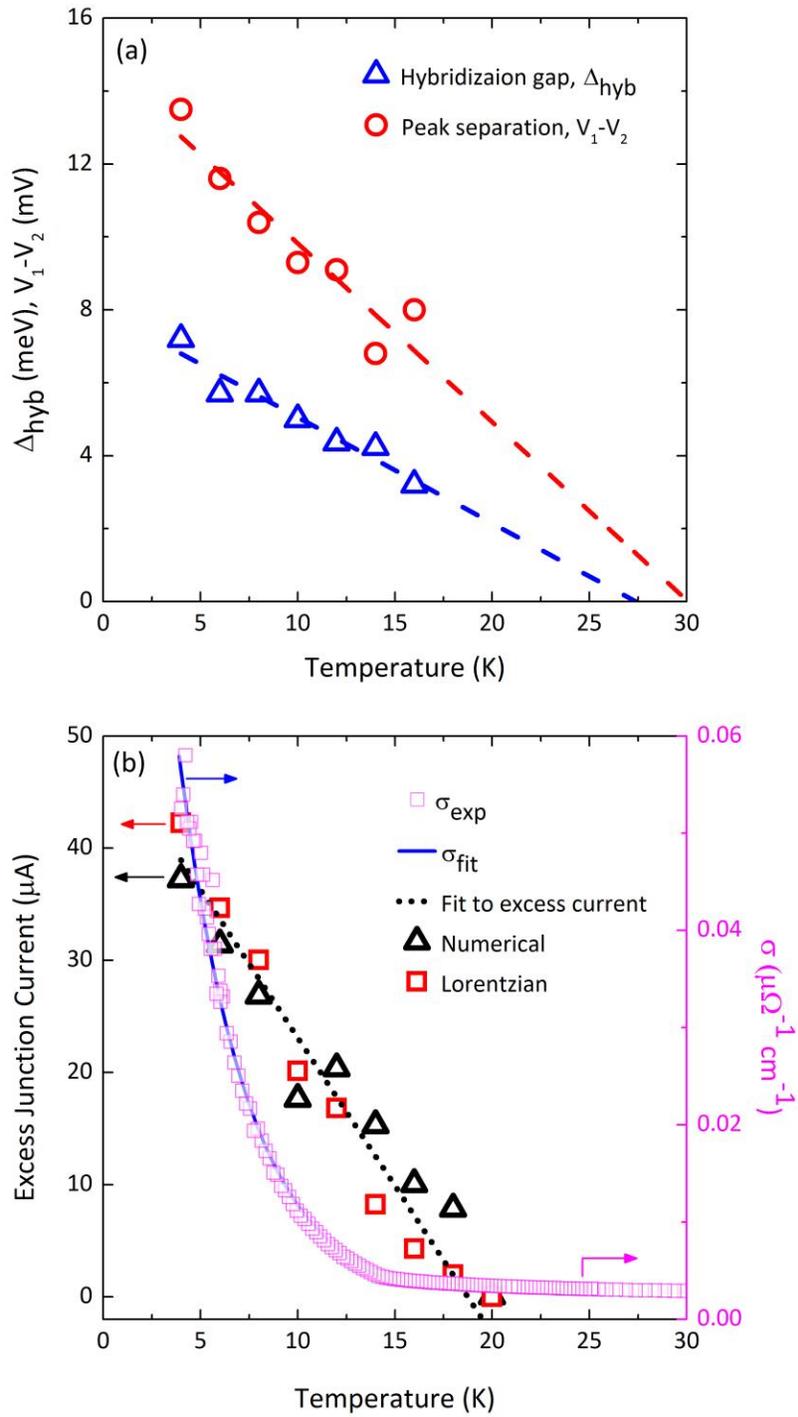

Figure 7

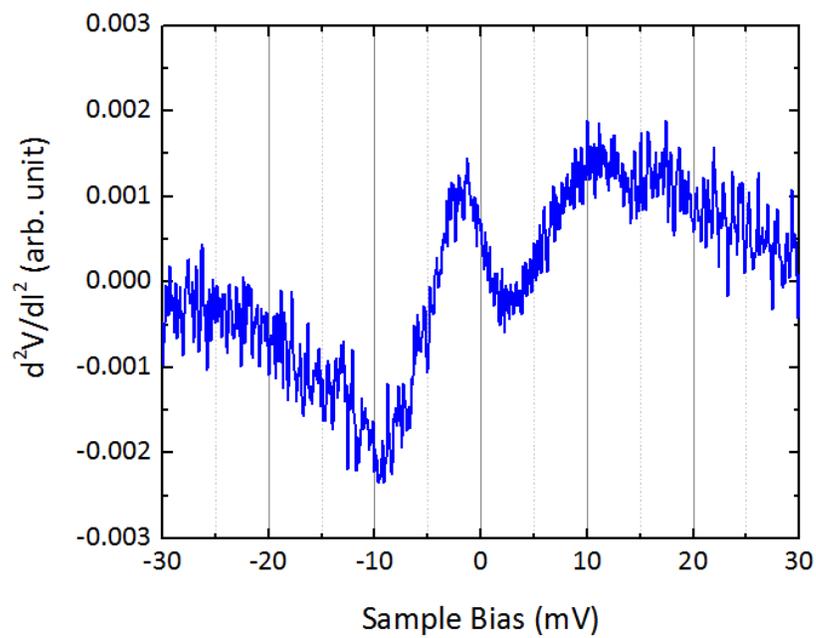

Figure 8